\documentclass[a4paper,aps,prd,reprint,superscriptaddress,nofootinbib,notitlepage,showpacs,showkeys,twocolumn,floatfix]{revtex4-1}
\usepackage{graphicx}
\usepackage{amssymb,amsmath,amsfonts}

\usepackage{dcolumn}
\usepackage{bm}
\usepackage{color}
\usepackage{dsfont}
\usepackage{mathrsfs}
\usepackage{multirow}

\newcommand \beq{\begin{eqnarray}}
\newcommand \eeq{\end{eqnarray}}

\newcommand{\MeV}{{\rm MeV}}
\providecommand{\e}[1]{\ensuremath{{\scriptscriptstyle E\negthinspace #1}}}

\begin{document}
\allowdisplaybreaks

\title{$O(4)$ $\phi^4$ model as an effective light meson theory: A lattice-continuum comparison}

\author{Gergely Mark{\'o}}
\email{marko@achilles.elte.hu}
\affiliation{Department of Theoretical Physics, E{\"o}tv{\"o}s University, P\'azm\'any P. s\'et\'any 1/A, H-1117 Budapest, Hungary.
}

\author{Zsolt Sz{\'e}p}
\email{szepzs@achilles.elte.hu}
\affiliation{MTA-ELTE Theoretical Physics Research Group, P\'azm\'any P. s\'et\'any 1/A, H-1117 Budapest, Hungary.}


\begin{abstract}
We investigate the possibility of using the four-dimensional $O(4)$ symmetric $\phi^4$ model as an effective theory for the sigma-pion system. We carry out lattice Monte Carlo simulations to establish the triviality bound in the case of explicitly broken symmetry and to compare it with results from continuum functional methods. In the case of a physical parametrization we find that triviality restricts the possible lattice spacings to a narrow range, and therefore cutoff independence in the effective theory sense is practically impossible for thermal quantities. We match the critical line in the space of bare couplings in the different approaches and compare vacuum physical quantities along the line of constant physics.
\end{abstract}


\maketitle

\section{Introduction}\label{sec:intro}

The $\phi^4$ scalar model with an internal $O(4)$ symmetry in 4 space-time dimensions has been long used as a model for spontaneous chiral symmetry breaking \cite{Lee72}. The direction of the symmetry breaking is associated with the sigma meson, while the pions are the Goldstone bosons emerging as a result of the spontaneous symmetry breaking. It is also widely known, that as a field theory it is trivial, it has no finite ultraviolet limit with non-zero coupling strength \cite{Luscher:1988uq}. Although this property is still discussed (see e.g. \cite{Shrock:2017zuk}) we accept it as a fact and investigate what is the bound set by triviality to the quantitative applicability of the model. Based on a calculation carried out by L\"uscher and Weisz (LW) in the same model applied to the Higgs particle \cite{Luscher:1988gc} one can estimate the lowest lattice spacing that can be reached in a parametrization adjusted to light mesons. This turns out to be $a_{\rm min}^{\rm LW} = 0.40(4)$~fm, which corresponds to a maximal cutoff in momentum representation to a few times $500$~MeV. This foreshadows that a scaling region of physical quantities as a function of $a$ on the lattice is unlikely to be found without getting too close to the triviality bound, and therefore cutoff independence, even in the effective theory sense is not feasible.

The above estimate was derived in a specific renormalization scheme for the case without explicit symmetry breaking. It is interesting to see to what extent it changes when, compared to \cite{Luscher:1988gc}, a different renormalization scheme is employed in the case when the pions are massive. At the same time, experience shows \cite{Marko:2013lxa} that the use of continuum functional methods is less restricted in the shadow of triviality and can retain some predictivity. To study this in more detail we use two continuum methods: the functional renormalization group (FRG) \cite{Wetterich:1992yh, Berges:2000ew} in the local potential approximation (LPA) and the two-loop and ${\cal O}(g_0^2)$ truncations of the two-particle irreducible approach (2PI) \cite{Cornwall:1974vz,Berges:2005hc,Marko:2013lxa}. Treating the model as a cutoff theory, we solve it using the same bare couplings as in the lattice version along the line of constant physics (LCP). Then, to compare the values of physical quantities, we need the relation between the lattice spacing $a$ and the cutoff $\Lambda$. This is determined by matching the critical line of the model at zero temperature  with the one determined by L\"uscher and Weisz in \cite{Luscher:1988uq} using the hopping parameter expansion.

The paper is structured as follows. In Sec.~\ref{sec:gen} we introduce notations for the model and summarize the details of the lattice simulations. In Sec.~\ref{sec:lcp} we define the LCP and describe how the triviality bound is obtained. We also discuss the immediate consequences of the value of the minimal lattice spacing. In Sec.~\ref{sec:comparison} we compare the lattice results with those obtained in the continuum approximations, and finally in Sec.~\ref{sec:conc} we summarize our findings.

\section{Generalities}\label{sec:gen}
We discuss the $O(N)$ symmetric, Euclidean $\phi^4$ model specifically for $N=4$, described in terms of bare quantities denoted by the subscript $0$ by the continuum action (omitting the obvious $x\equiv (t,\vec x)$ dependencies)
\beq
\nonumber
&&S_{\rm C}=\\
\nonumber &&\int d^4 x \left\{ \frac{1}{2}\partial_\mu \vec\phi_0\partial_\mu \vec\phi_0 + \frac{m_0^2}{2}\vec\phi_0\vec\phi_0 + \frac{g_0}{24}\left(\vec\phi_0\vec\phi_0\right)^2 - \vec H_0\vec\phi_0\right\}\,,\\
\label{eq:Scont}
\eeq
where $\vec\phi_0$ is the $N=4$ component field, $m_0$ is the mass, and $g_0$ is the quartic coupling. In the explicit symmetry breaking term the external field $\vec H_0$ is chosen to point in the direction of the first component of the scalar field with a length of $H_0$, independent of $x$. Discretization on a periodic, four-dimensional cubic lattice consisting of $N_T\times N_S^3$ sites (using a forward derivative), and rewriting in terms of the hopping parameter $\kappa$ leads to the well known lattice action \cite{Luscher:1988uq}:
\beq
\nonumber
S_{\rm L} = &&\sum_x \Bigg\{\vec\varphi\vec\varphi - 2\kappa\sum_{\hat\mu=1}^4\vec\varphi(x)\vec\varphi(x+a\hat\mu) + \\
&&\lambda\left(\vec\varphi\vec\varphi-1\right)^2 - \lambda - \vec h\vec\varphi\Bigg\}\,,
\eeq
where $a$ is the lattice spacing and $\hat\mu$ is the usual four-dimensional unit vector. The connections between the continuum and the lattice parameters are
\beq
a\vec\phi_0 &=& \sqrt{2\kappa}\vec\varphi\,,\\
g_0 &=& \frac{6\lambda}{\kappa^2}\,,\label{Eq:g0-lambda-kappa}\\
a^3\vec H_0&=&\frac{\vec h}{\sqrt{2\kappa}}\,,\\
a^2m_0^2&=&\frac{1-2\lambda}{\kappa}-8\,.\label{Eq:m02-lambda-kappa}
\eeq

We use Monte Carlo integration with importance sampling to evaluate path integrals. Configuration generation is done by using a poor man's heat bath algorithm, in which each site is updated using ten metropolis steps before its neighbors are updated in order to make the new field value at the chosen site practically independent of its initial value. Between two heat bath sweeps we also include two overrelaxation sweeps in order to sample a much larger part of the phase space using the same number of configurations.

\section{Line of constant physics}\label{sec:lcp}

\subsection{Observables defining the LCP}
\begin{center}
\begin{figure}
\includegraphics[width=0.45\textwidth]{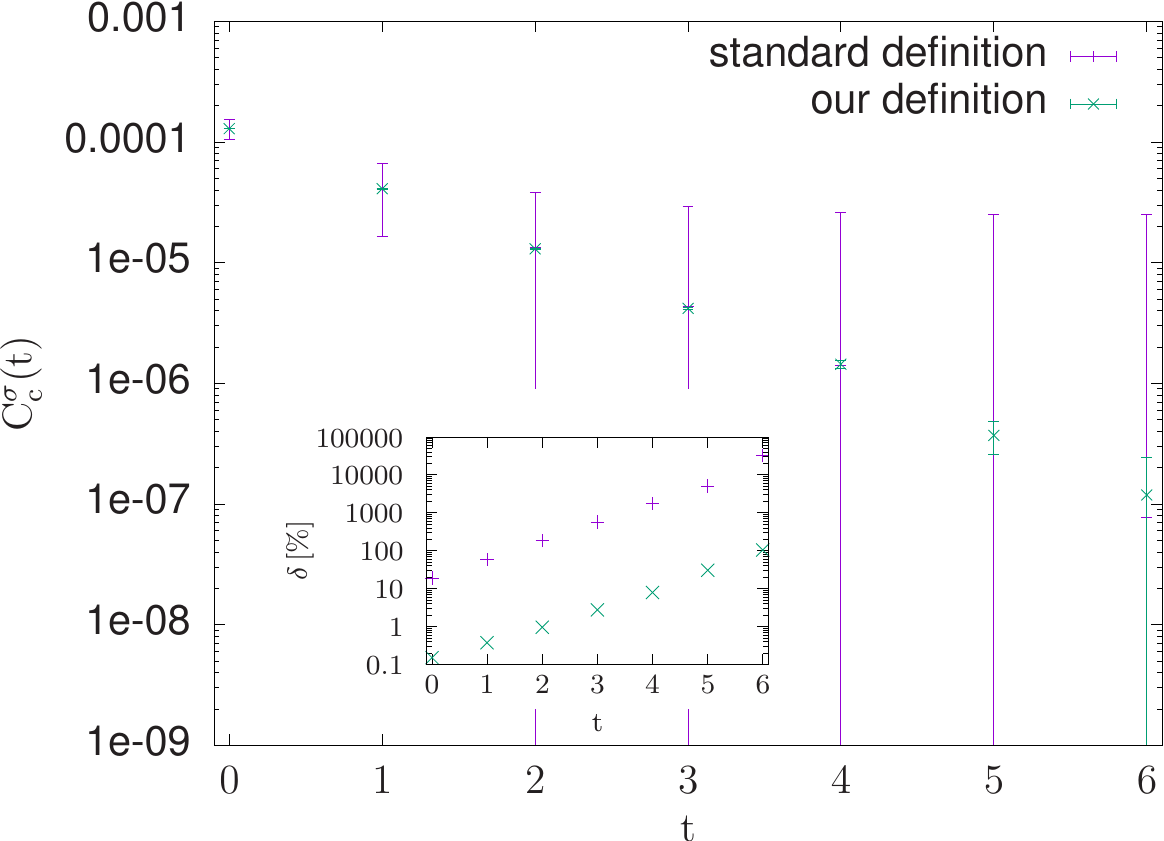}
\caption{An example of the two definitions of the connected sigma correlator. The purple points were obtained using the definition in \eqref{eq:ConnCorr}, while the green points were obtained using \eqref{eq:ConnCorrBetter}. The two definitions differ in a constant, but here they are shifted on top of each other for better comparison. \label{fig:CorrErr}}
\end{figure}
\end{center}

The explicitly broken $O(4)$ symmetric $\phi^4$ model has three parameters: the hopping parameter $\kappa$, the quartic coupling $\lambda$ and the external field $h$. In order to define a continuum limit\footnote{Even though a true continuum limit is not possible due to the triviality of the model, we follow the standard procedure which would allow to define it if it existed.} we give two physical prescriptions, which restrict our parameter space to the LCP, along which the lattice spacing $a$ tends to zero in physical units, at least in principle. The two prescriptions are
\begin{subequations}
\beq
\frac{m_\sigma a}{\bar\phi_R a}&=&\frac{300\,\MeV}{93\,\MeV}\approx3.226\,,\label{eq:lcp1}\\
\frac{m_\pi a}{\bar\phi_R a}&=&\frac{138\,\MeV}{93\,\MeV}\approx1.484\,,\label{eq:lcp2}
\eeq
\end{subequations}
where $m_{\sigma,\pi}$ are the respective pole masses and $\bar\phi_R$ is the expectation value (denoted by the bar) of the $\sigma$ component of the renormalized field, which takes the role of the pion decay constant in the linear sigma model (LSM). We choose a lower sigma mass (300~MeV) than what is generally agreed upon ($\approx 450$~MeV) \cite{Pelaez:2015qba}. Our choice is limited on the one hand by the fact that higher sigma masses are barely reachable in approximate continuum solutions of the LSM \cite{Patkos:2002xb,Andersen:2004ae} and on the other hand by the fact that we want to retain the kinematic possibility of the $\sigma\to2\pi$ decay.\\
To obtain the pole masses we measure time slice correlators. Let us define a time slice as
\beq
\vec s(t) = \frac{1}{N_S^3}\sum_{\vec x} \vec\varphi (t, \vec x)\,,
\eeq
and then
\beq
C_{ij}(t) = \frac{1}{N_T}\sum_{\tau} s_i(t)s_j(t-\tau)
\eeq
is the time slice correlator matrix for one configuration. The ensemble average of $C_{ij}(t)$ is the time slice correlator. By our choice of $\vec h$ the $\sigma$ direction is $i=1$; therefore $C_\sigma(t)\equiv C_{11}(t)$ is dominated by $m_\sigma$, while $C_{ii}(t)\,,i\neq 1,$ are all dominated by $m_\pi$. We do a least squares fit using the function\footnote{Contributions of excited states, if they exist, are invisible within our precision: the remnant obtained by subtracting the fitted form from the data is consistent with zero.}
\beq
f(t)=A+B\left(\exp(-mt)+\exp(-m(N_T-t))\right)\,,
\eeq
with parameters $A\,,B$ and $m$ to $C_\sigma(t)$ as well as to the average of the three pion directions\footnote{Averaging over the three pion directions lowers the statistical error of $m_\pi$ compared to $m_\sigma$.}
\beq
C_\pi(t) = \frac{1}{3}\sum_{i = 2}^4 C_{ii}(t)\,.
\eeq
The average and error of the fit parameters and in particular the masses are obtained by a jackknife analysis. The fit is carried out on each jackknife sample, leaving out the $t=0$ point of the correlator from the data in order to lower the distortions caused by possible higher excitations.\\
In the case of the sigma mass, one must take care of the disconnected part of the correlator. The connected part of the correlator is
\beq
\langle C_{\sigma,{\rm c}}(t)\rangle = \langle C_{\sigma}(t)\rangle - \langle M_1 \rangle^2\,,
\label{eq:ConnCorr}
\eeq
where $M_1$ is the first component of the average field over one configuration
\beq
\vec M=\frac{1}{N_TN_S^3}\sum_x \vec\varphi(x)\,.
\eeq
To subtract the correlated errors from the connected sigma correlator, instead of \eqref{eq:ConnCorr} we use another prescription (the two definitions differ only in a constant),
\beq
\langle C_{\sigma,{\rm c}}(t)\rangle = \langle C_{\sigma}(t) - M_1^2\rangle\,.
\label{eq:ConnCorrBetter}
\eeq
The definition \eqref{eq:ConnCorr} has a bad signal to noise ratio due to correlated errors which are canceled in \eqref{eq:ConnCorrBetter} leading to a better signal. We show the reduction of error achieved by using the definition in \eqref{eq:ConnCorrBetter} in Fig.~\ref{fig:CorrErr}.
The measurement of $\bar\phi_R$ goes as follows. The ensemble average of the first component (the sigma direction) of $\vec M$ is
\beq
\langle M_1 \rangle = \frac{a\bar\phi_0}{\sqrt{2\kappa}}\,,
\eeq
where the $0$ index on the right-hand side denotes that $\phi_0$ is a bare fields; that is, wave function renormalization is still needed. Then the renormalized vacuum expectation value is
\beq
\bar\phi_R = \frac{\bar\phi_0}{\sqrt{Z}}\,.
\eeq
We obtain $Z$ by prescribing the value of the zero-momentum inverse pion propagator to be the pion pole mass:
\beq
G_{R,\pi}^{-1}(p=0) \overset{!}{=} m_\pi^2\,.
\eeq
Through a Ward identity \cite{ZinnJustin:2002ru} the inverse two-point function can be rewritten as
\beq
\label{eq:WI}
G_{R,\pi}^{-1}(p=0) = \frac{H_R}{\bar\phi_R} = Z\frac{H_0}{\bar\phi_0}\,,
\eeq
which, in terms of lattice quantities and combined with the renormalization prescription, leads to
\beq
\label{eq:Z}
Z=2\kappa m_\pi^2\langle M_1\rangle h^{-1}\,.
\eeq
The value of $Z$ is slowly changing between 0.74 and 0.8 along the LCP in the measured range of $a$.

\subsection{Determining the LCP \label{sec:det_lcp}}

\begin{figure}
\begin{center}
\includegraphics[width=0.45\textwidth]{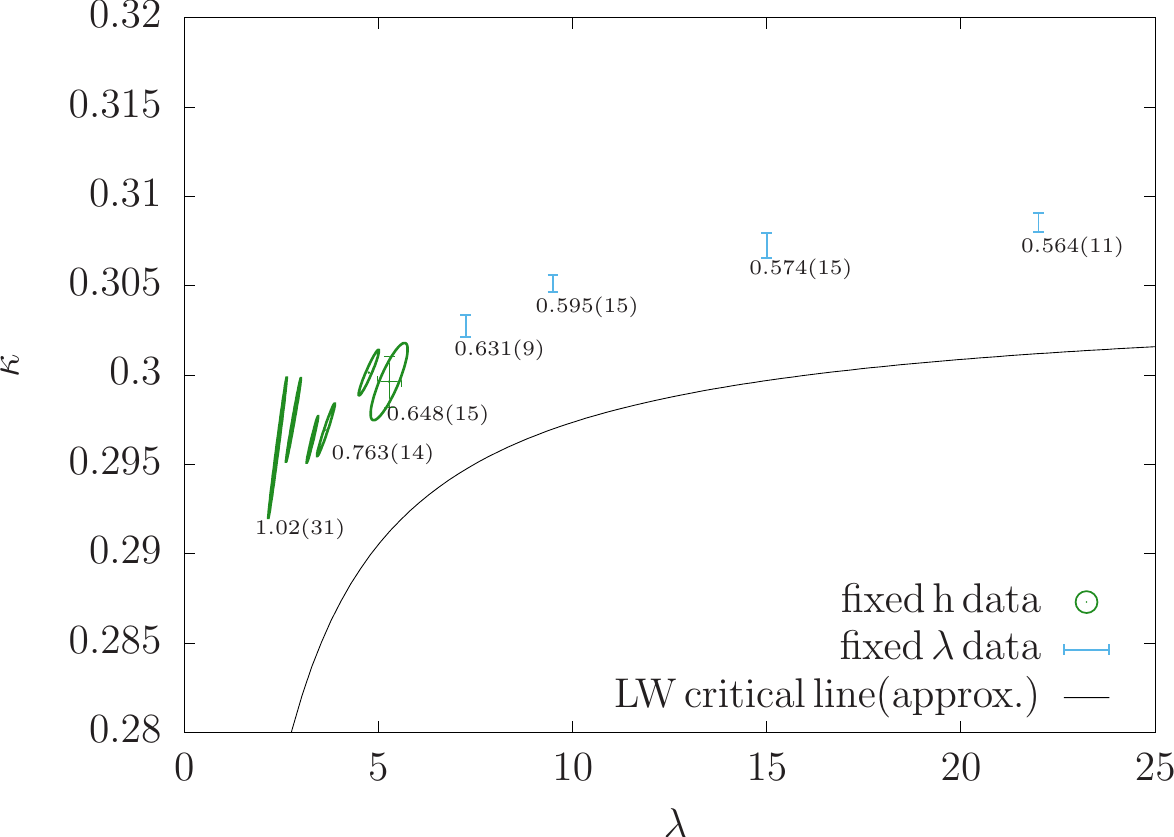}
\caption{Points of the LCP in the $\lambda$-$\kappa$ plane labeled by the value of the lattice spacing (values not shown are listed in Table~\ref{Tab:data}). A simple interpolating approximation of the critical line based on the results of L\"uscher and Weisz \cite{Luscher:1988uq} is also shown for orientation. The ``fixed $h$'' and ``fixed $\lambda$'' indicate that a grid taken in a constant $h$ or $\lambda$ plane, respectively, was used to measure the ratios appearing in \eqref{eq:lcp1} and \eqref{eq:lcp2}. The interpolation procedure used to find the LCP points is described in the text and detailed in Appendix~\ref{app:lcp_search}. The ellipse represents the 68.3\% confidence level associated with the bootstrap configurations and the error bar indicates the standard deviation of $\kappa$ (and in one case also of $\lambda$) over the bootstrap sample.}
\label{fig:lcp}
\end{center}
\end{figure}

To obtain a point of the LCP curve, we fix the value of one of the parameters (usually $h$, but in the region where triviality strongly influences the LCP we fix $\lambda$) and measure the ratios of observables appearing in \eqref{eq:lcp1} and \eqref{eq:lcp2} on an appropriate grid in the plane of the remaining two parameters, $\kappa-\lambda$ and $\kappa-h$ planes respectively. The physical values of ratios define contour lines and the LCP is obtained as the intersection of two contour lines, each belonging to one surface. Points of the contour lines are obtained by linear interpolation between grid points and are fitted with parabolas. The intersection of the two parabolas is one point of the LCP corresponding to the $h$ or $\lambda$ where the grid was defined. The error is estimated by a bootstrap resampling using $10^4$ samples. A detailed description of this procedure is relegated to Appendix~\ref{app:lcp_search}. The original ratios of observables were obtained using $16\times16^3$ lattices with $10^5$ field configurations. While $16\times m_\pi a > 6$ even for the smallest lattice spacing, we tested volume independence by checking at the last LCP point that the observable ratios of \eqref{eq:lcp1} and \eqref{eq:lcp2} only change within errors when $20\times20^3$ and $24\times24^3$ lattices are used.

With the method outlined above, we obtain the points of the LCP shown in Fig.~\ref{fig:lcp}. The conversion of $a$ to physical units is done using
\beq
a\bar\phi_R  = a_{\rm phys} \times 93 \MeV = a_{\rm phys} \times (93/197.327)\,{\rm fm}^{-1}\,.\quad
\eeq

\begin{figure}
\begin{center}
\includegraphics[width=0.45\textwidth]{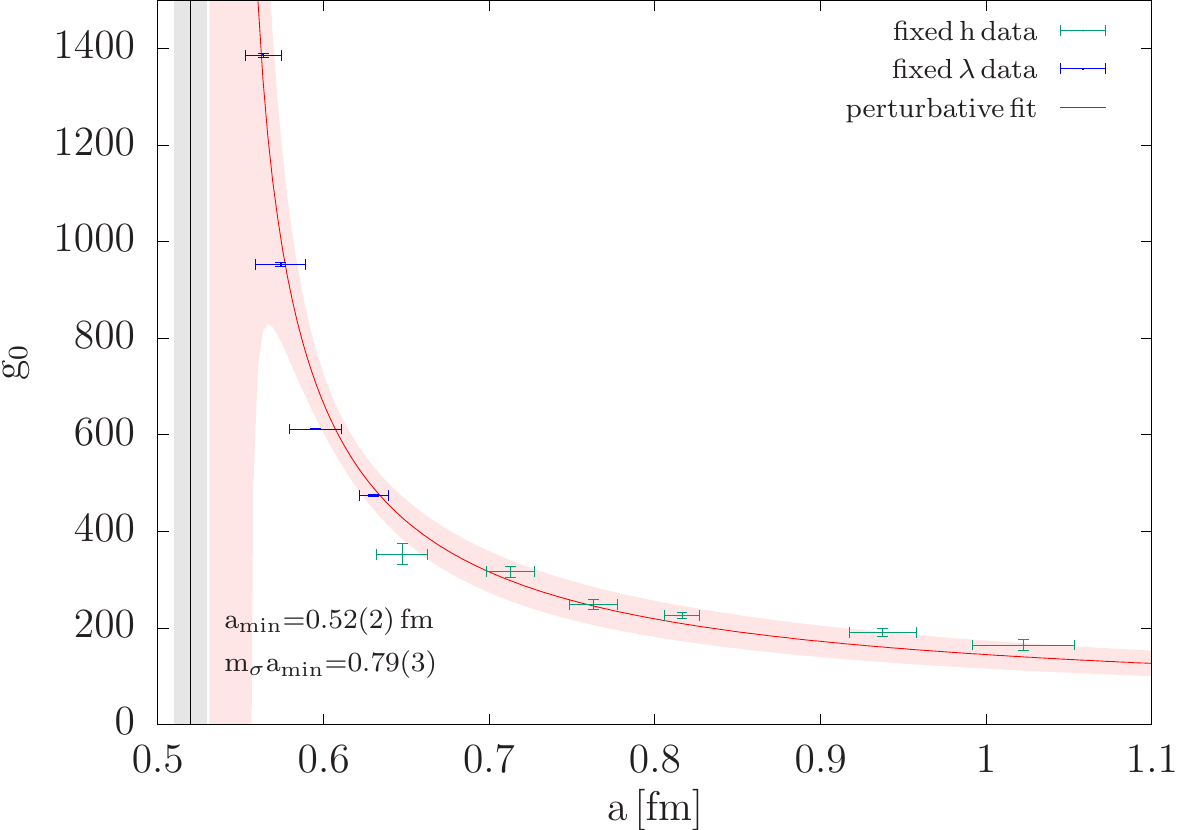}
\caption{The original bare self-coupling defined in \eqref{eq:Scont} as a function of the lattice spacing. The second order perturbative $\beta$ function \eqref{eq:betaFunc} is also shown (red line, standard deviation shaded), with the parameters $g_1$ and $a_1$ being fitted to the data. Using this functional form we can also estimate the triviality bound $a_{\rm min}.$ The error was obtained using bootstrap resampling which also had samples having their poles around $a=0.55$~fm, causing the standard deviation to grow enormously in that region.}
\label{fig:trivi}
\end{center}
\end{figure}

In Fig.~\ref{fig:lcp} we see that the LCP follows the critical line in the $\lambda-\kappa$ plane. In a theory with a proper continuum limit the LCP should run into the critical line at least at infinite coupling. While in our case $h\neq0$ for all LCP points, $h$ tends to zero along the line as it should to approach criticality. Triviality appears here by seeing that even at $\lambda\to\infty$ the LCP does not converge to the critical line, meaning that $a$ remains finite. This means that the bare $\phi^4$ coupling $g_0$ must have a pole as a function of $a$, at the minimal value of the lattice spacing. The results for $g_0$ are shown in Fig.~\ref{fig:trivi} and are in compliance with the generally accepted view on the triviality of the $\phi^4$ model. Fitting the data shown in Fig.~\ref{fig:trivi} with the second order perturbative $\beta$ function
\beq
g_0^{-1}(a) = g_1^{-1}-(\beta_1+\beta_2 g_1)\log(a_1/a)\,,
\label{eq:betaFunc}
\eeq
where
\beq
\beta_1 = \frac{1}{3}\frac{N+8}{8\pi^2}\,,\quad\beta_2=-\frac{1}{3}\frac{3N+14}{(16\pi^2)^2}
\eeq
are the standard $\beta$-function coefficients \cite{Smit:2002ug}, we estimate $a_{\rm min}=0.52(2)$~fm. This leads to an estimate for the minimal value of the lattice $\sigma$ mass, $am_\sigma=0.79(3)$.

The result for the minimal lattice spacing can be compared to the one which can be given based on \cite{Luscher:1988gc}. In the renormalization scheme of L\"uscher and Weisz
\beq
\label{eq:LWgR}
g_R=\frac{3m_R^2}{v_R^2}\,,
\eeq
where $g_R$ is the renormalized quartic coupling, $m_R$ is the renormalized mass, which we identify with the sigma mass for the sake of the estimate, and $v_R\equiv \bar\phi_R$ takes the value of the pion decay constant as in our case, although the $Z$ factor, which we do not need here, is defined differently. In \cite{Luscher:1988gc} the renormalization trajectories are described, and taking the $\lambda\to\infty$ limit in them yields a relation between $m_Ra_{\rm min}$ and $g_R$:
\beq
\label{eq:LWamin}
\log\big(m_R a_{\rm min}\big) = \frac{1}{\beta_1 g_R}+\frac{\beta_2}{\beta_1^2}\log(\beta_1 g_R)-1.9(1)\,,
\eeq
where the number $1.9(1)$ is the result of a numerical calculation at a high order of the hopping parameter expansion. Plugging $m_R = 300$~MeV and $v_R = 93$~MeV into \eqref{eq:LWgR} and \eqref{eq:LWamin} yields $a_{\rm min}^{\rm LW}=0.4$~fm already mentioned in Sec.~\ref{sec:intro}. We see that our result $a_{\rm min}=0.52(3)$ is even more restrictive.

An important implication of the largeness of $a_{\rm min}$ is that on the lattice the maximal temperature that can be simulated is $T = (N_t\cdot a_{\rm min})^{-1}|_{N_t=1} \approx 420$~MeV. Furthermore if one is interested in a ``continuum limit'' in the effective theory sense the feasible temperature range is definitely below $50$~MeV. This limits the comparison of the continuum methods practically to vacuum quantities.

\section{Comparison with 2PI and FRG results} \label{sec:comparison}

Since according to the previous section, a comparison between lattice and continuum physical quantities is not feasible at a finite temperature, we remain at $T=0$ and using continuum functional methods we determine the masses along the LCP shown in Fig.~\ref{fig:lcp}. In order to compare to a lattice result determined at a fixed lattice spacing, we need to treat the continuum version of the model as a cutoff theory. Hence we need the relation between the continuum cutoff $\Lambda$ and the lattice spacing $a$, {i.e.} we need $c=\Lambda a$. This relation was studied in \cite{Brahm:1994zv,Besjes} where the conversion factor $c\approx4.9$ was calculated analytically for the four-dimensional (4D) hypercubic lattice and obtained also by fitting the perturbative continuum result (using \eqref{Eq:g0-lambda-kappa} and \eqref{Eq:m02-lambda-kappa})
\beq
\label{Eq:m20c-pert}
-\frac{m^2_{0,c}}{\Lambda^2} = \frac{N+2}{6}\left[\frac{g_0}{(4\pi)^2} - \frac{2}{3}\frac{g_0^2}{(4\pi)^4} + \mathcal{O}(g_0^4)\right]
\eeq
to the critical line $m^2_{0,c}(g_0)$ obtained by L\"uscher and Weisz in \cite{Luscher:1988uq}. The above equation comes from the condition of the vanishing curvature mass at the vanishing field value at second order in the perturbation theory. We can see in Fig.~\ref{fig:crit-line} that at $\mathcal{O}(g_0^2)$ it reproduces the LW critical line only at small values of the coupling. One expects that this behavior changes if one uses a more sophisticated approximation.

In Fig.~\ref{fig:crit-line} we also show the LCP points corresponding to the ``fixed h'' data points of Fig.~\ref{fig:lcp}. While doing the transformation of the bootstrap data from the $\lambda-\kappa$ plane to the $g_0-m_0^2$ plane, we have also performed a principal component analysis of the correlation matrix. As indicated in Fig.~\ref{fig:crit-line}, the bootstrap sample in the $g_0-m_0^2$ plane is much more elongated in one direction than it was in the $\lambda-\kappa$ plane. Based on the standard deviations of $g_0$ and $m_0^2$ over the bootstrap sample, shown for the rightmost LCP point, one could be incorrectly led to think that the LCP is compatible with the critical line. This is not the case, of course, as the bootstrap sample is well separated from the critical line, as was the case already in the $\lambda-\kappa$ plane.

\subsection{The critical line $m^2_{0,c}(g_0)$ in the 2PI framework \label{ss:2PI-crit_line}}

\begin{center}
\begin{figure}
\includegraphics[width=0.45\textwidth]{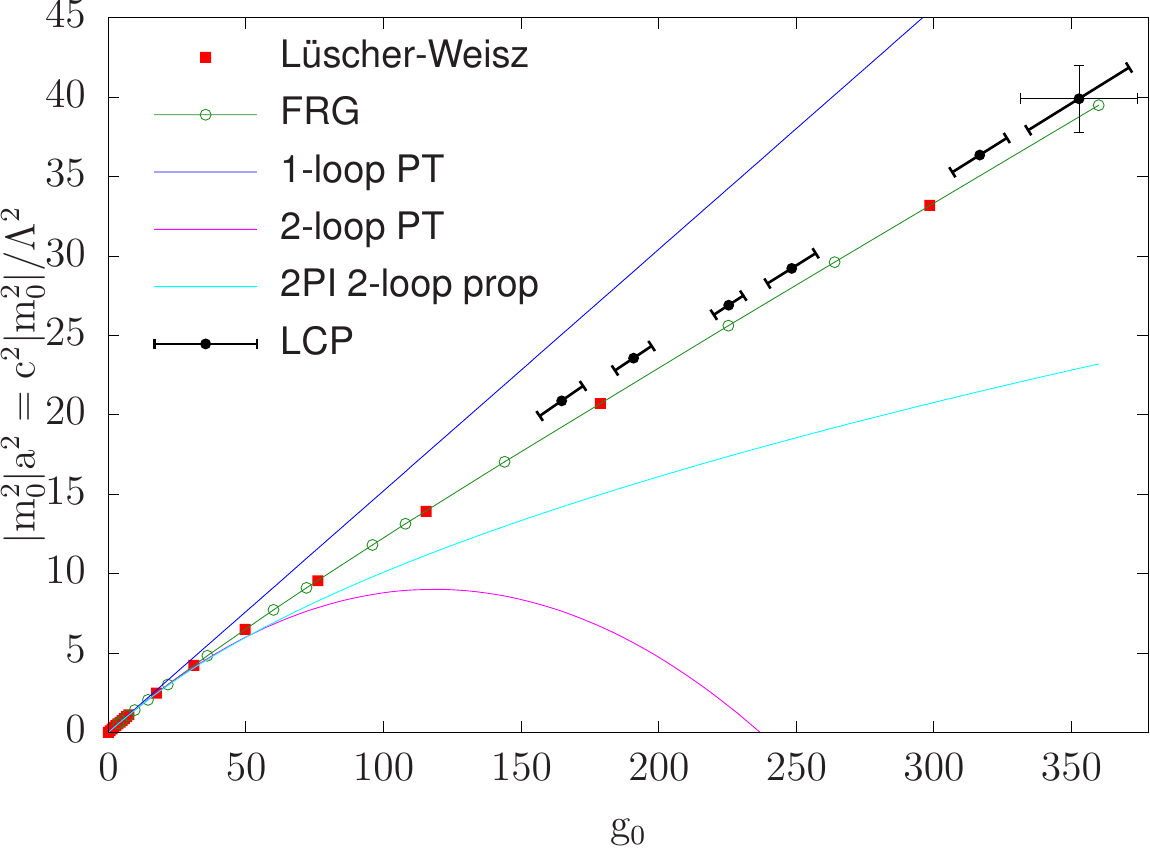}
\caption{The critical line determined in the continuum theory using various approximations as compared to that obtained with the hopping parameter expansion to 14th order. In the perturbation theory (PT) and the 2PI approach $c=4.9$, while in the FRG study $c=6.923$. The points of the LCP corresponding to the fixed $h$ data points shown in Fig.~\ref{fig:lcp} with ellipses are also presented. With the exception of the rightmost point of the LCP, the error bar corresponds to the standard deviation (square root of the larger eigenvalue of the covariance matrix) along the major axis of the bootstrap sample (the minor axes are too small to be seen on this scale). For the rightmost point we also show the standard deviation of $g_0$ and $|m_0^2|a^2$ over the bootstrap sample.}
\label{fig:crit-line}
\end{figure}
\end{center}

In the 2PI framework, the curvature mass at the vanishing field value is given at the $\mathcal{O}(g_0^2)$ level of truncation of the effective action by $\hat M^2_{\phi=0}=\bar M^2_{\phi=0}(K=0)$, where the gap mass satisfies the self-consistent equation \cite{Marko:2015gpa}
\beq
\label{Eq:gap_eq_order_g2}
\bar M^2_{\phi=0}(K)=m_0^2+\frac{N+2}{6}\left[g_0{\cal T}[\bar G]-\frac{g^2_0}{3}{\cal S}[\bar G](K)\right].\ \
\eeq
The tadpole and setting-sun integrals involve the propagator $\bar G(K)=1/(K^2+\bar M^2_{\phi=0}(K)).$ The critical line is determined from the condition of the vanishing curvature mass: $\hat M^2_{\phi=0}=\bar M^2_{\phi=0}(K=0)=0$.

The nontrivial momentum dependence makes \eqref{Eq:gap_eq_order_g2} rather hard to solve, however, the solution $\bar M^2_{\phi=0}(K=0)$ can be approximated by using a localized propagator with momentum independent mass gap $\bar M^2=m_0^2+\frac{N+2}{6}g_0{\cal T}[\bar G_{\rm loc}]$, where $\bar G_{\rm loc}(K)=1/(K^2+\bar M^2).$ This approximation corresponds in fact to the two-loop 2PI truncation. In this approximation, the tadpole can be explicitly computed with a 4D cutoff $\Lambda$ and the condition of the vanishing curvature mass can be written as
\begin{subequations}
\beq
\tilde M^2&=&\frac{N+2}{18}g_0^2\tilde {\cal S}(\tilde M^2),
\label{Eq:SS-gap}
\\
\frac{m_{0,c}^2}{\Lambda^2}&=&\tilde M^2-\frac{(N+2)g_0}{96\pi^2}\left[1-\tilde M^2\ln\left(1+\tilde M^{-2}\right)\right],\qquad
\label{Eq:m02-2loop_approx}
\eeq
\end{subequations}
where $\tilde{\cal S}(\tilde M^2)$ is a perturbative setting-sun integral at vanishing external momentum and we used the tilde for a quantity scaled by appropriate powers of $\Lambda$.

For a given $g_0$, one then solves \eqref{Eq:SS-gap} for $\tilde M^2$ and using this solution one has $m_{0,c}^2$ from \eqref{Eq:m02-2loop_approx}. The critical line obtained in this way is shown in Fig.~\ref{fig:crit-line}. It still deviates from the LW curve, but remains closer to it in a wider range of the coupling than the $\mathcal{O}(g_0^2)$ perturbative curve. We mention that \eqref{Eq:m20c-pert} can be obtained by first using \eqref{Eq:SS-gap} in the first term on the right-hand side of \eqref{Eq:m02-2loop_approx} and then taking the $\tilde M^2\to 0$ limit, in which $\tilde{\cal S}(\tilde M^2)\to 2/(4\pi)^4$.

Now let us discuss the determination of the critical line by solving \eqref{Eq:gap_eq_order_g2} without further approximation. $m^2_{0,c}(g_0)$ could be obtained in principle by approaching it from the symmetric phase: fixing $g_0$, the equation is solved for increasing values of $|m_0^2|$ and $m^2_{0,c}(g_0)$ is obtained by extrapolating the determined values of $\bar M^2_{\phi=0}(K=0)$ to zero. As detailed in Appendix~\ref{app:SSgapSol}, Eq.~\eqref{Eq:gap_eq_order_g2} is solved by treating the setting-sun ${\cal S}(K)$ as a double convolution: a convolution of the propagator with a bubble integral, where the latter is itself a convolution of two propagators. It turned out that the solution to \eqref{Eq:gap_eq_order_g2} is lost for a value of $m^2_0$ where $\bar M^2_{\phi=0}(K=0)$ is nonzero (see Fig.~\ref{Fig:gap-sol}). This loss of solution, which seems to be a feature of the $\mathcal{O}(g_0^2)$ 2PI gap equation, and was investigated in detail in \cite{Marko:2015gpa}, prevents us from the direct determination of the critical line at this order of the 2PI truncation scheme and furthermore from a comparison along the LCP.

As Fig.~\ref{fig:crit-line} shows, the simpler two-loop approximation indeed has a critical line determined by Eqs.~\eqref{Eq:SS-gap} and \eqref{Eq:m02-2loop_approx}. Nevertheless, a loss of solution can also happen in this approximation in the broken phase (that is at $\phi\neq0$) depending on the parameters \cite{Marko:2015gpa}. We found that the usual iterative procedure to solve the broken phase two-loop equations (which were written down and solved as detailed in \cite{Marko:2013lxa} with little modifications to accommodate for the use of nonrenormalized equations and approximating the $T\to0$ limit numerically) breaks down close to the critical line in comparison to where the points of the LCP are; therefore in the LCP points no solution exists and no comparison can be made. We checked that this loss of solution persists in the even simpler localized two-loop approximation which we detailed in \cite{Marko:2015gpa}. We conclude that in the considered approximations the 2PI formalism cannot be compared to the lattice LCP results.

\subsection{Determination of observables using the FRG method}

Another functional method from which one can calculate curvature masses along the LCP is the functional renormalization group method. The flow equation describing the evolution of the scale-dependent average action $\Gamma_k$ from the ultraviolet (UV) scale $k=\Lambda$, where the microscopic theory is defined through the bare action, down to the deep infrared (IR), where the usual quantum effective action is obtained in the $k\to 0$ limit, is \cite{Wetterich:1992yh}
\beq
\partial_k \Gamma_k[\phi] = \frac{1}{2}{\rm Tr}\left[\partial_k R_k\left(\frac{\delta^2 \Gamma_k[\phi]}{\delta\phi_i\delta\phi_j}+R_k\delta_{ij}\right)^{-1}\right],
\label{Eq:WE}
\eeq
where $R_k$ is a regulator function, that is, in momentum space it suppresses the IR modes, while $\partial_k R_k$ regulates the integral in the UV. In the LPA the Ansatz
\beq
\Gamma_k[\phi]=\int d^dx\left(\frac{1}{2}\big(\partial_x\phi_i\big)^2+U_k(\rho)\right),
\eeq
is used, where $\rho=\vec{\phi}^2/2$ is $O(N)$ invariant and it is customary to choose the LPA-optimized regulator \cite{Litim:2001up} $R_k(q)=(k^2-q^2)\Theta(k^2-q^2)$ ($q$ is the Euclidean four-momentum). Then, using $\displaystyle\frac{\partial^2 U_k(\rho)}{\partial\phi_i\partial\phi_j}=U_k'(\rho)(P^{\rm L}_{ij}+P^{\rm T}_{ij})+2\rho U_k''(\rho) P^{\rm L}_{ij}$ with $P^{\rm L/T}$ being the longitudinal/transverse projectors, the integral can be performed and, at zero temperature and $d=4$, one obtains
\beq
\partial_k U_k(\rho)=\frac{k^5}{32 \pi^2}\left(\frac{N-1}{k^2+\hat M^2_{\rm T}(k)}+\frac{1}{k^2+\hat M^2_{\rm L}(k)}\right),\
\label{Eq:U-flow}
\eeq
where $\hat M^2_{\rm T}(k)=U_k'(\rho)$ and $\hat M^2_{\rm L}(k)=M^2_{\rm T}(k)+2\rho U_k''(\rho).$ This equation is solved numerically by integrating it down to $k=0$ (in practice to some $k_{\rm end}>0$, due to the flattening of the potential) starting at scale $k=\Lambda,$ where the initial condition for the potential is given in terms of the couplings $m_0^2$ and $g_0$ as $U_{k=\Lambda}(\rho)=m_0^2\rho+g_0\rho^2/6.$

In the so-called grid method $U_k(\rho)$ is discretized using $N_\rho$ grid points so that \eqref{Eq:U-flow} transforms into a system of $N_\rho$ coupled ordinary differential equations. We solve this system using the Runge-Kutta-Fehlberg algorithm with adaptive step-size control provided by the GNU Scientific Library (GSL)\cite{GSL}. We work in units of the cutoff, denoting with tilde a quantity scaled with the cutoff, and we choose $N_\rho=5000$ equidistant values of $\tilde\rho=\rho/\Lambda$ in the range between 0 to $\tilde\rho_{\rm max}=0.026$. The flow was stopped at $\tilde k_{\rm end}=1.28\times 10^{-2}$ where all the monitored quantities became practically constants. At each point of the grid the first and second order derivatives of the potential are calculated with ${\cal O}(\Delta\tilde \rho^4)$ finite difference formulas. The minimum of the potential is obtained with spline interpolation, while the curvature masses at the minimum are obtained fitting a sixth order polynomial to the potential in an appropriate $\tilde\rho$ interval which has the minimum as its left end point.

\begin{center}
\begin{figure}
\includegraphics[width=0.45\textwidth]{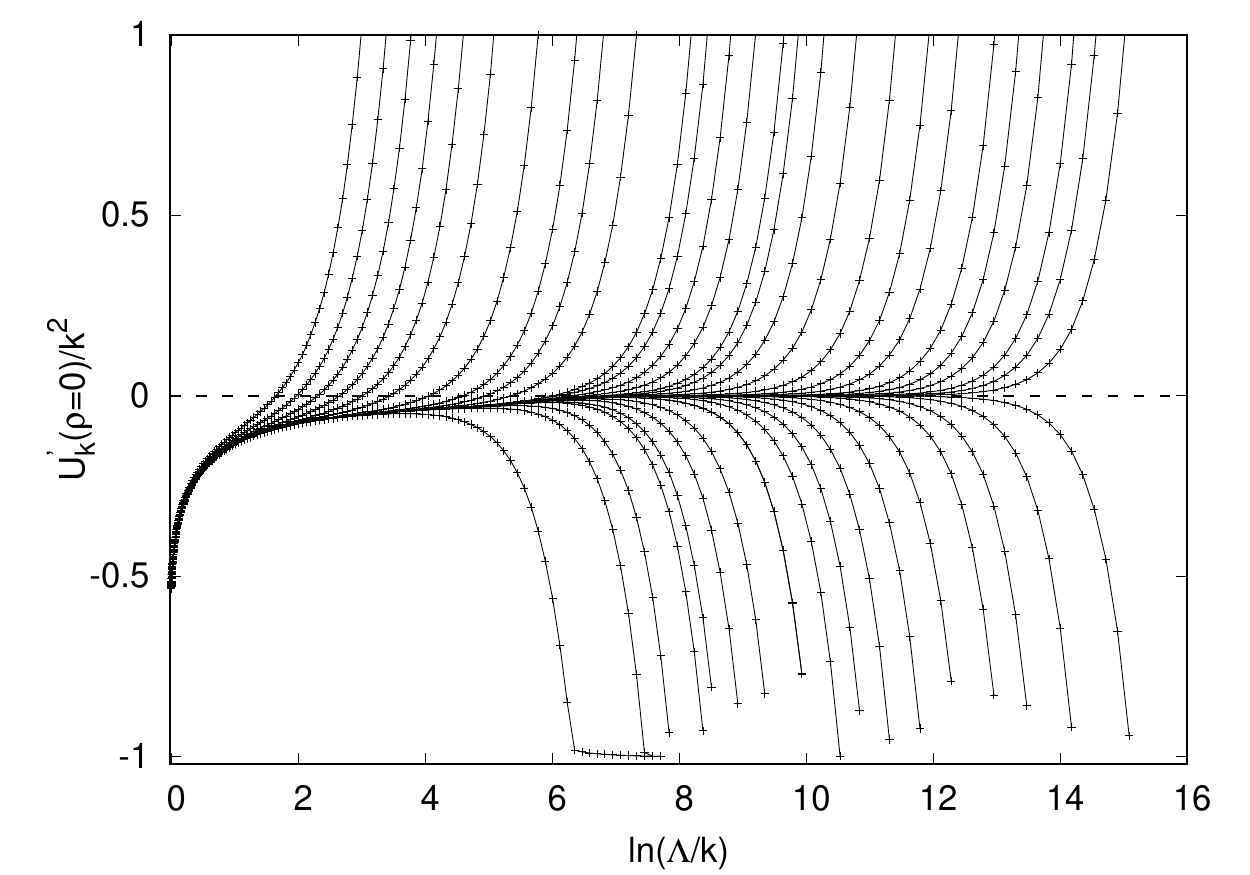}
\caption{Determining $\tilde m_{0,c}^2$ at $g_0=225.36$ by dichotomy. For $|\tilde m_0^2|>|\tilde m_{0,c}^2|$ we are in the broken symmetry phase and $U'_k(0)/k^2\to -1$ as $k\to0$, while for $|\tilde m_0^2|<|\tilde m_{0,c}^2|$ we are in the symmetric phase and $U'_k(0)/k^2\to \infty$ as $k\to0$. The closer $|\tilde m_0^2|$ is to the critical value, the larger is $\ln(\Lambda/k)$ at which a curve steeply goes upwards or downwards.}
\label{fig:caillol}
\end{figure}
\end{center}

The transverse and longitudinal curvature masses are obtained at $k_{\rm end}$ as $\hat M_{\rm T}=\sqrt{U_k'(\bar\rho)}$ and  $\hat M_{\rm L}=(\hat M^2_{\rm T}+2\bar\rho U_k''(\bar\rho))^{1/2}$, where $\bar\rho=\bar\phi^2/2$ is the minimum of the potential. In the LPA they can be regarded as approximations to the pole masses due to the simplicity of the Euclidean propagator. In order to compare $\hat M_{\rm T/L}$ along the LCP with the values $M_\sigma=300$~MeV and $M_\pi=138$~MeV which in the lattice simulation are constant along the line, we need to know what is the relation between the cutoff scale $\Lambda$ and the lattice spacing $a$. This relation is obtained by matching the critical curve $\tilde m_{0,c}^2(g_0)$ determined in the FRG case to the one obtained by L\"uscher and Weisz in \cite{Luscher:1988uq} using the hopping parameter expansion to 14th order. We determine $\tilde m_{0,c}^2(g_0)$ working at fixed $g_0$ and using dichotomy on $\tilde m_{0}^2$, as shown in Fig.~\ref{fig:caillol}, where the quantity that distinguishes between the broken and symmetric phases is $U'_k(0)/k^2$.

\begin{table*}
  \caption{Field and curvature mass values in units of the cutoff at the minimum of the potential of the LCP shown in Figs.~\ref{fig:lcp} and \ref{fig:crit-line}. The points are denoted by $P_i$ with $i\in1,\dots,10$ in increasing order from left to right of the LCP. For the first six (fixed $h$) points the values comes from the direct numerical solution of \eqref{Eq:U-flow}, while for the last four (fixed $\lambda$) points the values come from the solution obtained using the expansion \eqref{Eq:kifejt} with an extrapolation to $N_g=\infty$. \label{Tab:data}}
\centering
\begin{tabular}{ c c c c c c c c c c c c } \hline\hline\\[-8pt]
Point & $m_0^2 a^2$ & $g_0$ & $H_0 a^3$ & $a$[fm] & $\bar\phi/\Lambda$ & $\hat M_{\rm T}/\Lambda$ & $\hat M_{\rm L}/\Lambda$ & $\Lambda\mathrm{[MeV]}=\frac{6.923}{a\mathrm{[fm]}}$ & $\bar\phi$[MeV]  & $\hat M_{\rm T}$[MeV] & $\hat M_{\rm L}$[MeV] \\[1pt] \hline
P1 & -2.087\e{+1} & 1.648\e{+2} & 2.599\e{-1} & 1.022\e{+0} & 6.471\e{-2} & 1.100\e{-1} & 2.712\e{-1} & 1335.804 & 86.438 & 146.983 & 362.300 \\
P2 & -2.356\e{+1} & 1.909\e{+2} & 1.944\e{-1} & 9.375\e{-1} & 5.934\e{-2} & 9.938\e{-2} & 2.465\e{-1} & 1457.094 & 86.459 & 144.808 & 359.174 \\
P3 & -2.690\e{+1} & 2.254\e{+2} & 1.298\e{-1} & 8.164\e{-1} & 5.182\e{-2} & 8.691\e{-2} & 2.113\e{-1} & 1673.228 & 86.701 & 145.425 & 353.486 \\
P4 & -2.922\e{+1} & 2.484\e{+2} & 1.038\e{-1} & 7.632\e{-1} & 4.849\e{-2} & 8.033\e{-2} & 1.953\e{-1} & 1789.874 & 86.784 & 143.775 & 349.481 \\
P5 & -3.635\e{+1} & 3.167\e{+2} & 8.389\e{-2} & 7.130\e{-1} & 4.503\e{-2} & 7.493\e{-2} & 1.820\e{-1} & 1916.000 & 86.280 & 143.566 & 348.727 \\
P6 & -3.990\e{+1} & 3.528\e{+2} & 6.458\e{-2} & 6.475\e{-1} & 4.125\e{-2} & 6.869\e{-2} & 1.640\e{-1} & 2109.794 & 87.026 & 144.931 & 345.914 \\
\hline
P7 & -5.259\e{+1} & 4.746\e{+2} & 5.463\e{-2} & 6.305\e{-1} & 3.870\e{-2} & 6.517\e{-2} & 1.533\e{-1} & 2166.640 & 84 & 141 & 332 \\
P8 & -6.699\e{+1} & 6.122\e{+2} & 5.152\e{-2} & 5.952\e{-1} & 3.753\e{-2} & 6.417\e{-2} & 1.497\e{-1} & 2295.396 & 86 & 147 & 344 \\
P9 & -1.024\e{+2} & 9.534\e{+2} & 4.431\e{-2} & 5.742\e{-1} & 3.392\e{-2} & 6.257\e{-2} & 1.446\e{-1} & 2379.292 & 85 & 149 & 344 \\
P10 & -1.474\e{+2} & 1.387\e{+3} & 4.203\e{-2} & 5.637\e{-1} & 3.532\e{-2} & 6.095\e{-2} & 1.558\e{-1} & 2423.379 & 86 & 148 & 377 \\
\hline
\hline
\end{tabular}
\end{table*}

Once we match $\tilde m_{0,c}^2$ and $ m_{0,c}^2 a^2$ at some value of $g_0$, finding the relation
\beq
a\Lambda\approx6.923,
\label{Eq:c_FRG}
\eeq
the entire critical curve determined using FRG agrees with the one obtained by L\"uscher and Weisz, as shown in Fig.~\ref{fig:crit-line}. The very good agreement of the two critical curves is in line with the findings of Ref.~\cite{Caillol:2012dt}, where it was reported that in the one component $\phi^4$ model the critical line, obtained in the LPA with the Litim regulator and lattice discretization, compares well with the one determined with Monte Carlo simulations.

Having obtained the relation between $a$ and $\Lambda$, we can now solve the flow equation \eqref{Eq:U-flow} and determine the curvature masses for the fixed $h$ data points of Fig.~\ref{fig:lcp}. The results shown in the first four rows of Table~\ref{Tab:data} in units of the cutoff can be used in two ways. In the first case, shown in the last four columns of Table~\ref{Tab:data}, one can determine for each point of the LCP the value of the cutoff from the lattice spacing using \eqref{Eq:c_FRG}. Then $\bar\phi$ is smaller than $f_\pi=93$~MeV by $\sim8\%,$ $\hat M_{\rm T}$ is $5\%-8\%$ larger than $M_\pi=138$~MeV, while $\hat M_{\rm L}$ is $15\%-20\%$ larger than $M_\sigma=300$~MeV. In the second case one can require $\bar\phi$ to be $f_\pi$. In this case, due to the larger value of the cutoff, one finds that $\hat M_{\rm L}$ is $22\%-30\%$ larger than the sigma values used to determine the LCP, while $\hat M_{\rm T}$ is larger by around $10\%$ than the pion mass. The deviation from the lattice results decreases for smaller $a$.

\subsubsection{Solution of a modified flow equation for $m_0^2<-\Lambda^2<0$}

With the chosen quartic potential at the initial value of the scale, $k=\Lambda,$ the flow equation \eqref{Eq:U-flow} cannot be solved for $m_0^2<-\Lambda^2<0$ due to a singularity in the equation. One could either change the initial condition by including higher order, perturbatively nonrenormalizable terms in the potential or, as we do it here following \cite{Barnafoldi:2016tkd}, try to circumvent the problem by modifying the flow equation expanding in power series to some order $N_g$ the fractions appearing in the right-hand side of \eqref{Eq:U-flow},
\beq
\frac{1}{k^2+\hat M^2_{\rm L/T}}=\frac{1}{k^2+M_0^2}\,\frac{1}{1-\xi} \approx\frac{1}{k^2+M_0^2}\sum_{n=0}^{N_g}\xi^n,\quad
\label{Eq:kifejt}
\eeq
where $\xi=(M_0^2-\hat M^2_{\rm L/T}(k))/(k^2+M_0^2)$ with $M_0$ some large parameter, {\it i.e.} $M_0>\Lambda$, which for numerical reasons has to be chosen appropriately.\footnote{We use the same values of $M_0$ and $N_g$ for both fractions in \eqref{Eq:U-flow}.}

First, keeping the numerical framework used so far, that is changing only the right-hand side of \eqref{Eq:U-flow} according to \eqref{Eq:kifejt}, we tested the method in a case where a direct solution to \eqref{Eq:U-flow} exists and then we applied it for the fixed $\lambda$ data points of the LCP shown in Fig.~\ref{fig:lcp} (points P7-P10 in Table~\ref{Tab:data}). In the latter case the solution is regarded as an approximation to the solution of the original Wetterich equation \eqref{Eq:WE}, assumed to exist for an appropriate form of the effective action at scale $\Lambda$.

In the case of point P1, it turns out that in order to reproduce the available direct solution of \eqref{Eq:U-flow} with the expansion method, one has to go to rather high orders in the expansion. Also, for the method to work, the first and second derivatives of the potential at $\tilde\rho_{\rm max}$ had to be kept fixed as a function of $k$, however, the chosen values were practically arbitrary. We fixed the derivatives to their values calculated at $k=\Lambda$.

\begin{center}
\begin{figure}
\includegraphics[width=0.45\textwidth]{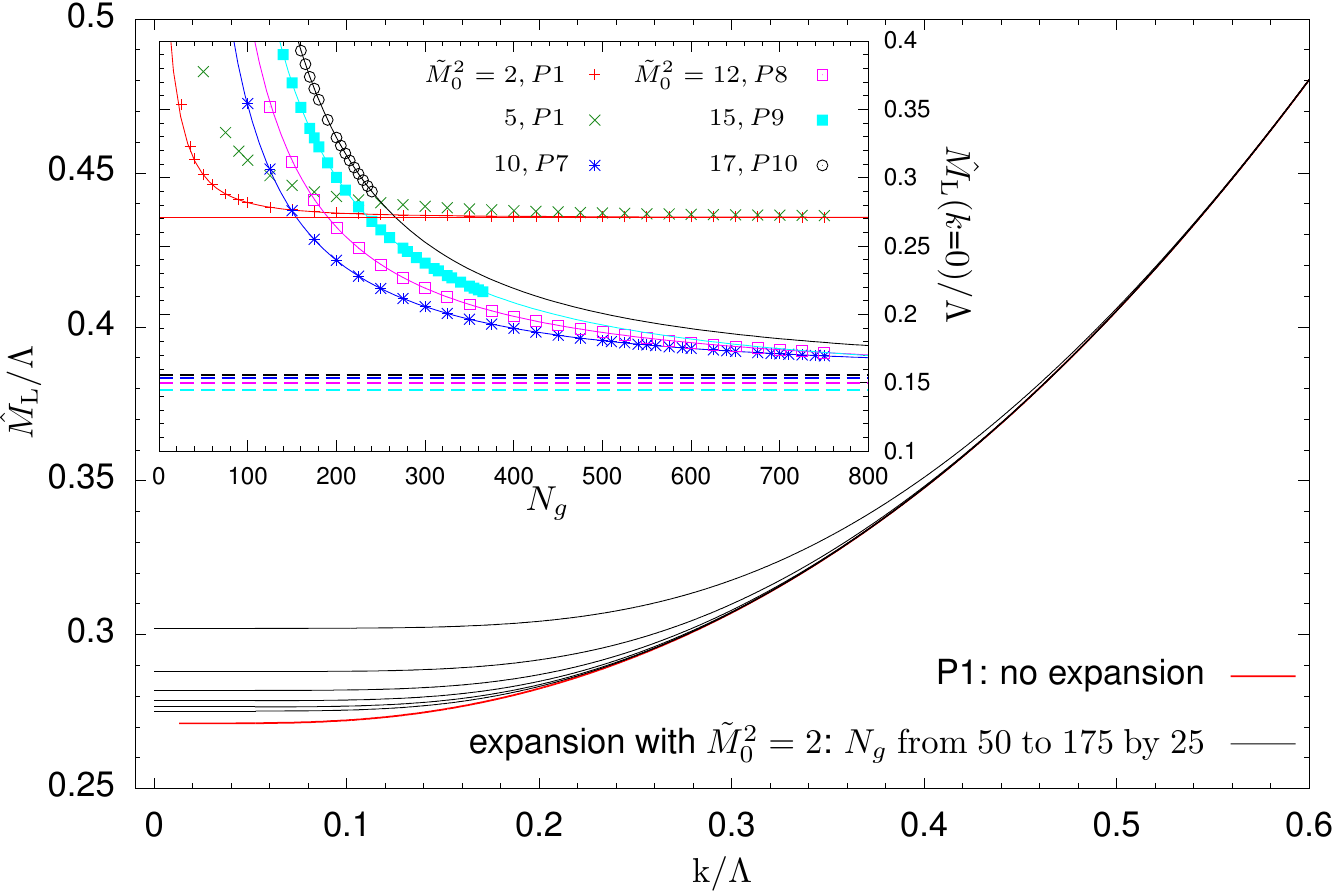}
\caption{Convergence properties of the solution to the modified flow equation for parameters corresponding to points of the LCP (see Table~\ref{Tab:data} for their labeling). The outset shows the scale dependence of the longitudinal curvature mass $\hat M_{\rm L}$ at various orders of the expansion for a point of the LCP where the solution of the original LPA flow equation \eqref{Eq:U-flow} is known. The inset shows the dependence of $\hat M_{\rm L}$ at $k_{\rm end}=0$ on the order of the expansion, and the influence of the expansion point $M_0^2$, also in cases when the original equation \eqref{Eq:U-flow} has no solution. The horizontal lines indicate the value at $N_g=\infty$ extracted from fits. }
\label{fig:FRG_cont}
\end{figure}
\end{center}

At a given order of the expansion the deviation from the direct result increases with $M_0.$ Among the studied quantities, $\hat M_{\rm L}$, presented in Fig.~\ref{fig:FRG_cont}, shows the slowest convergence rate with $N_g$ at a fixed value of $M_0$. For $\tilde M_0^2=2$ and $N_g=50$ the deviation from the direct result is around $10\%$. To estimate the result of the curvature masses and the minimum of the potential we fitted with $f(x)=a+b/(x-c)^d$ the data obtained at various $N_g$ with the expansion method. For P1 one can practically recover the direct results from a dataset obtained with up to $N_g\simeq 100$ terms in the expansion, but as $|m_0^2|/\Lambda^2$ increases we need larger $M_0^2$ and larger values of $N_g$ to maintain the quality of the fit. Eventually, numerical errors prevent us from going above a certain value of $N_g$. All these features are illustrated in Fig.~\ref{fig:FRG_cont} and the results obtained with the expansion values are given in the last four rows of Table~\ref{Tab:data}. Based on the variation of the extrapolated results on the fitting $N_g$-interval, one can estimate the error of $\hat M_{\rm L}$ to be $1\%-2\%$ for P7 and P8 and $5\%-10\%$ for P9 and P10. For the other two quantities the error of the extrapolation to $N_g=\infty$ is smaller.

\section{Conclusion}\label{sec:conc}
We studied the four component Euclidean $\phi^4$ model in four dimensions. In the presence of an explicit symmetry breaking term, we determined with Monte Carlo simulations the line of constant physics (LCP) in the bare parameter space of the model based on ratios involving the pion and sigma masses and the expectation value of the field. In this process we brought further evidence in support of the triviality of the model in a renormalization scheme which is different from the one usually used by the lattice community (see \cite{Siefert:2014ela} for a recent study).

Using the bare couplings of the LCP, we solved the model with two continuum functional methods (the 2PI formalism and the FRG method) in an attempt to compare the vacuum masses and expectation value obtained with these continuum methods to the corresponding input values of the lattice study of the model. The manifestation of triviality prevented us from a meaningful comparison of finite temperature quantities. It turned out that the comparison at $T=0$ can be done only with the FRG, since the 2PI is hindered by the loss of solution to the propagator equation. The needed relation between the lattice spacing and the cutoff, used in the lattice and continuum versions of the model, respectively, was obtained by matching the critical line of the parameter space determined originally by L\"uscher and Weisz using hopping parameter expansion.

\begin{acknowledgments}
The authors would like to thank Sz.~Bors\'anyi, S.~D.~Katz, G.~ Fej\H{o}s, A. Jakov\'ac and Cs.~T{\"o}r{\"o}k for helpful and interesting discussions. In the case of G.~M. this work is part of Project No. 121064 for which support was provided from the National Research, Development and Innovation Fund of Hungary, financed under the PD\_16 funding scheme.  Zs.~Sz. acknowledges support by the J\'anos B{\'o}lyai Research Scholarship of the Hungarian Academy of Sciences.
\end{acknowledgments}

\appendix

\section{Finding the LCP point and its error \label{app:lcp_search}}

We detail here the procedure used to find the points of the LCP. The seemingly overly complex procedure summarized in Sec.~\ref{sec:det_lcp} is needed because the observable ratios \eqref{eq:lcp1} and \eqref{eq:lcp2} change very similarly over the three-dimensional parameter space. Therefore, the search for the LCP becomes the mathematical problem of finding the intersection of two noisy surfaces which are almost parallel.

Our generic approach to find a point of the LCP is to fix one parameter and scan the remaining two for the pair of parameters where the observable ratios have both the prescribed physical values. For points of the LCP that are far from the triviality bound, that is $a_{\rm latt}$ is relatively large, we fix $h$ and scan the $\kappa$-$\lambda$ plane. As the lattice spacing becomes smaller, we change to fixing $\lambda$ and scan the $\kappa$-$h$ plane.


We discuss in what follows the case when we fix $h$ and search for the intersection point of two contour lines in the $\kappa$-$\lambda$ plane. The two curves are marked out by \eqref{eq:lcp1} and \eqref{eq:lcp2}, which are almost parallel locally, as mentioned earlier. We scan the plane by doing simulations in a set of $(\kappa,\lambda)$ pairs forming a square grid. In each point of the grid we independently measured the observable ratios ($R_1$ and $R_2$) with known errors ($\Delta_1$ and $\Delta_2$). To estimate the location of the LCP point in the plane we generate $10^4$ bootstrap configurations on the grid, where the value of both observable ratios at each grid point is randomly taken from independent normal distributions with means $R_{1,2}$ and standard deviations $\Delta_{1,2}$. Each bootstrap configuration can then be thought of as two surfaces $R_{1,2}(\kappa,\lambda)$ over the parameter plane sampled on the grid points.

For each bootstrap configuration we linearly interpolate the observable surfaces along the edges of the square grid. We locate the points which belong to the physical contour lines of the respective surfaces (physical values of $R_1$ and $R_2$). These two sets of points are then fitted by parabolas. The intersection of the two parabolas is one bootstrap realization of the LCP point. The bootstrap realizations outline the probability distribution of the LCP point in the $\kappa$-$\lambda$ plane, and hence its error can be estimated.

\begin{center}
\begin{figure}
\includegraphics[width=0.45\textwidth]{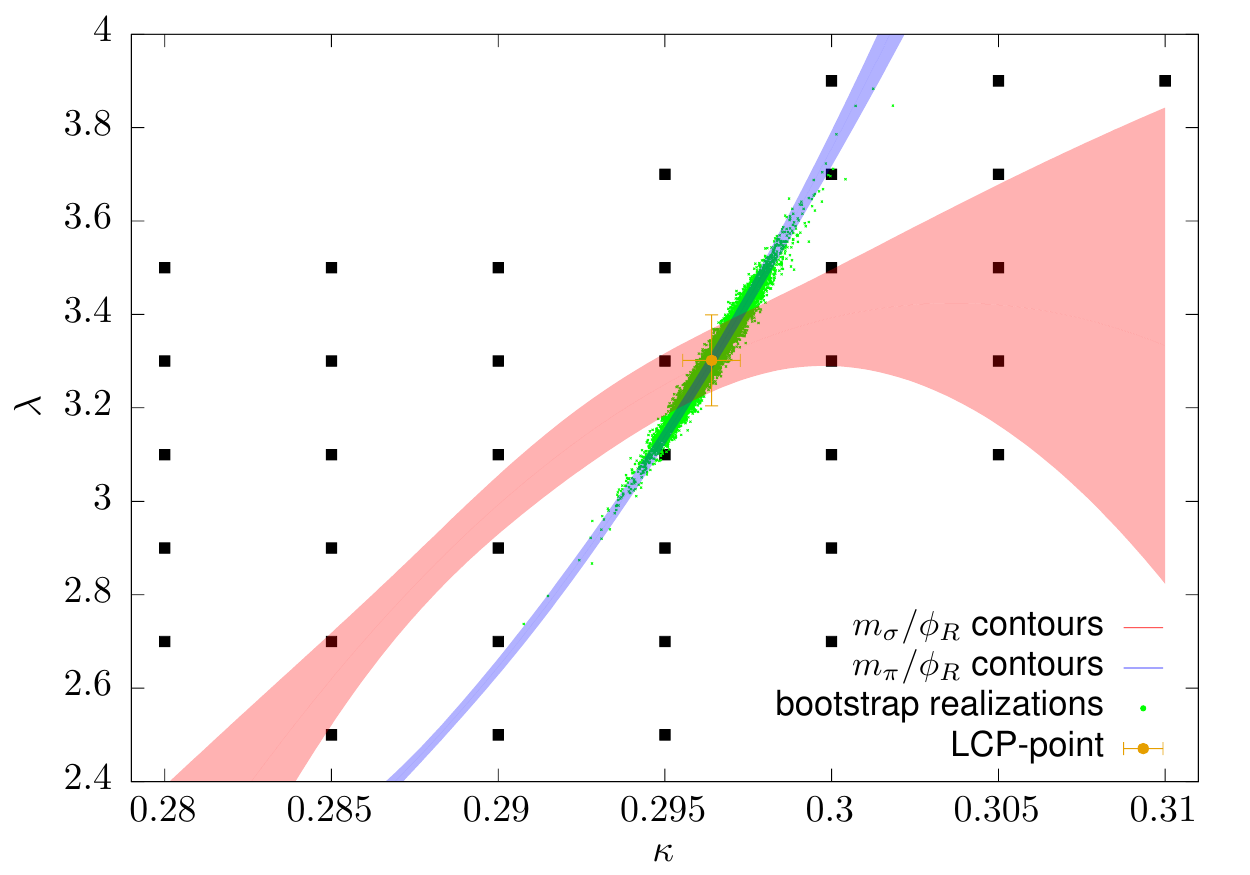}
\caption{Output of the LCP-point finding procedure at $h=0.1.$ The grid of black squares marks the parameter values where simulations were carried out, the bands show the average and error of the contour lines where the observables $R_1$ and $R_2$ take their physical values, the green dots are the bootstrap realizations of the LCP point and the final estimate for the LCP is the yellow blob, with error bars representing the standard deviation of $\kappa$ and $\lambda$.}
\label{fig:lcpbootsrap}
\end{figure}
\end{center}

The procedure is illustrated in Fig.~\ref{fig:lcpbootsrap} for $h = 0.1$ (P3 of Table \ref{Tab:data}), which shows the bootstrap realization of the LCP point and its estimated average value and error. The bands correspond to the averages and $1\sigma$ deviation of the parabolas approximating the physical contours. The distribution shows correlation between the $\kappa$ and $\lambda$ coordinates of the LCP point and it is elongated along one of the contour lines. This elongation is further enhanced when transformed to $m_0^2$ and $g_0$, where the distribution becomes practically one dimensional. As a result we show error bands in the direction of the major axis of the distribution for each LCP point in Fig.~\ref{fig:crit-line}.

By appropriately choosing the grid and including more points in it, as well as increasing the precision of the lattice simulations, one could in principle further confine the location of the LCP points.

\section{The  $T=0$ setting-sun integral as a Hankel transform \label{app:Hankel}}
Using the Fourier transform, the convolution of two momentum-dependent functions can be written as
\beq
C_{\rm 4d}[f_1,f_2](q) & = & \int_kf_1(k)f_2(q-k) \nonumber \\
& = & \int_x e^{-iq\cdot x} f_1(x) f_2(x),
\label{Eq:convol}
\eeq
where we used the shorthands $\int_k=\int\frac{d^4 k}{(2\pi)^4}$ and $\int_x = \int d^4x$. Working with spherical coordinates in 4D, the angular integral can be performed analytically by exploiting the rotation invariance. Choosing $q$ to point in the fourth direction, such that $q\cdot x = Q X \cos\theta_1$, where $Q=|q|$ and $X=|x|,$ and introducing $\tau=\sin\theta_1,$ the nontrivial part of the angular integration gives a Bessel function of the first kind,
\beq
\int_{-1}^{1}d\tau \sqrt{1-\tau^2} e^{-i X Q\tau} = \frac{\pi J_1(Q X)}{Q X},
\eeq
such that \eqref{Eq:convol} becomes
\beq
C_{\rm 4d}[f_1,f_2](Q)&=&\frac{4\pi^2}{Q}\int_0^\infty d X X J_1(Q X) A(X) \nonumber \\
&=:& \frac{4\pi^2}{Q}H_1[A](Q),
\label{Eq:HT}
\eeq
where we used the Hankel transform of order 1 of the function $A(X)=F_1(X) F_2(X)/X$, with $F_i(X)=X f_i(X)$.

A similar calculation shows that $F_i(X)=X f_i(X)$ appearing in \eqref{Eq:HT} can be given as the inverse Hankel transform (denoted in what follows by a tilde) of order 1 of $\tilde F_i(P)= P \tilde f_i(P)$ ($i=1,2$):
\beq
X f_i(X) &=& \frac{1}{4\pi^2}\int_0^\infty d P P J_1(X P) \tilde F_i(P) \nonumber\\
&=:& \frac{\tilde H_1[\tilde F_i](X)}{4\pi^2}.
\label{Eq:IHT}
\eeq
So, in terms of Hankel transforms, the convolution \eqref{Eq:convol} can be written as
\beq
C_{\rm 4d}[f_1,f_2](Q)=\frac{H_1\big[X^{-1}\tilde H_1[\tilde F_1](X) \tilde H_1[\tilde F_2](X)\big](Q)}{4\pi^2 Q}.\qquad
\eeq
Then, writing the momentum dependent setting-sun as a convolution of a propagator and a bubble integral $B[G](q-k)=\int_p G(p) G(q-k-p)$, both the bubble integral and the setting-sun integral $S[G](q)=\int_k G(k) B(q-k)$ can be written in terms of Hankel transforms. The discrete version of the Hankel transform (DHT) and its inverse are implemented in the GSL package \cite{GSL}. In the discretized case it is understood that all momenta and combinations of momenta are cut by the cutoff $\Lambda.$ Working in units of the cutoff and using $3\times 2^{10}$ sampling points the smallest momentum on the grid is $|\tilde k|_{\rm min}=3.96\times 10^{-4}.$

\section{On the solution of \eqref{Eq:gap_eq_order_g2} \label{app:SSgapSol}}

As mentioned in Sec.~\ref{ss:2PI-crit_line}, the setting-sun ${\cal S}[G](K)$ can be regarded as a double convolution. The convolution integral can be computed using the Fourier transform, which for rotational invariant functions leads at $T=0$ to the use of the Hankel transform, as detailed in Appendix \ref{app:Hankel}. As discussed in Sec.~V.E.1 of \cite{Marko:2012wc}, calculating a convolution using discrete Fourier transform is not accurate if the function does not decrease fast enough in the UV. We expect this behavior in the case of the Hankel transform as well. Although to determine the critical line we are interested in an IR quantity, namely $\bar M^2_{\phi=0}(K=0),$ since a momentum integral is involved in its calculation, the discretization error in the UV will influence this quantity. To check the method that uses the Hankel transform, we also computed on a nonuniform momentum grid \footnote{We use $N_K=128-256$ values of momenta: $K_i=K_{\rm min}+(\Lambda-K_{\rm min})\,(i/(N_K-1))^{2.5}$, $i=0,\dots, N_K-1$ with $\tilde K_{\rm min}=3.96\times 10^{-4}$.} the convolution as a double integral using Eq.~(A1) of \cite{Fejos:2011zq}. The inner integral in that expression is calculated numerically after a Tanh-Sinh transformation (see Appendix A of \cite{Reinosa:2011cs} for details) using the splined $\bar M^2(K)$ in the propagator. The bubble integral is calculated in this way on a grid, then splined and used for the calculation of ${\cal S}(K).$

\begin{figure}
\begin{center}
\includegraphics[width=0.45\textwidth]{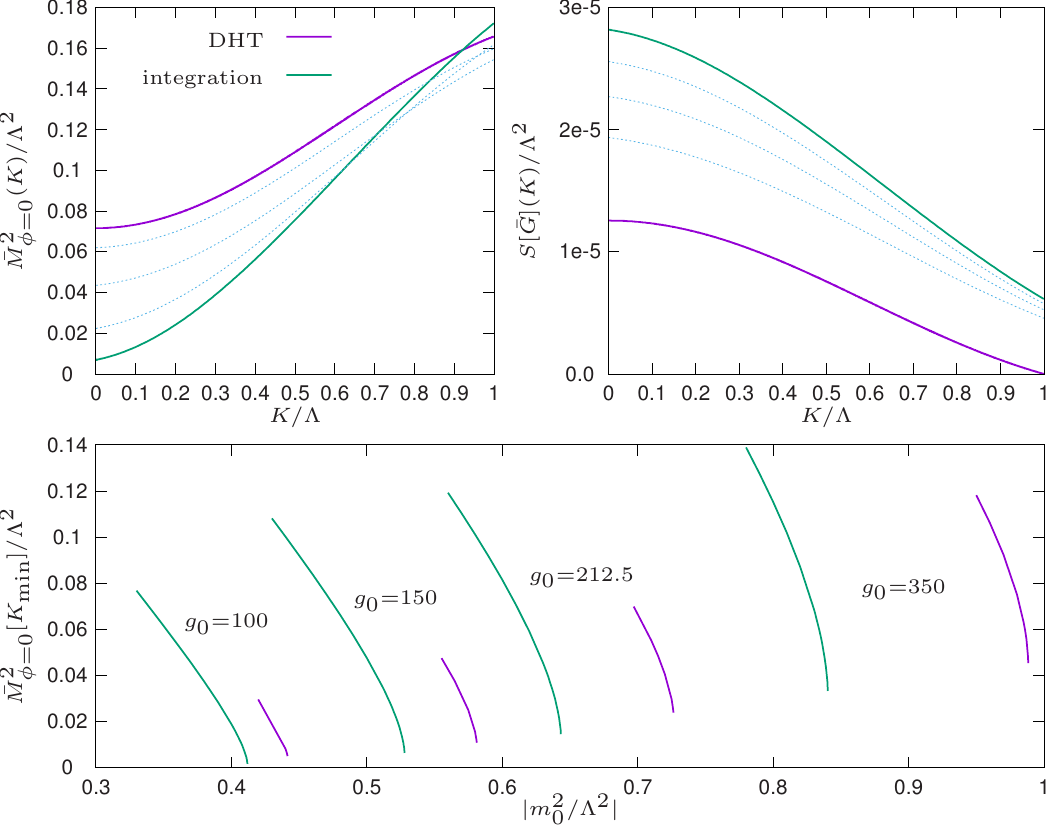}
\caption{Comparison of the solution of the gap equation \eqref{Eq:gap_eq_order_g2} obtained using the discrete Hankel transformation (purple lines) or numerical integration (green lines) to compute convolutions. In both top panels, where $g_0=150$ and $m_0^2/\Lambda^2=-0.5275$, the dotted blue lines correspond to the 1st, 4th and 26th iterations (from purple to green in order) if one initiates the numerical integration method from the solution obtained using DHT. Top left: both the functional form and the infrared limit of the momentum dependent gap mass differs significantly. Top right: the difference seen originates mainly from the different UV behavior of the setting-sun diagram shown here. Bottom: the infrared limit of the solutions of the gap equation as a function of $m_0^2$ for several fixed $g_0$ values. These curves should be extrapolated to $\bar M^2_{\phi=0}(K_{\rm min})=0$ in order to find the critical line, however a loss of solution at some finite $\bar M^2_{\phi=0}(K_{\rm min})$ prohibits doing so.}
\label{Fig:gap-sol}
\end{center}
\end{figure}

The iterative solution of \eqref{Eq:gap_eq_order_g2} obtained for $g_0=150$ and $m_0^2/\Lambda^2=-0.5275$ using underrelaxation method \cite{Berges:2004hn} with parameter $\alpha=0.1$ is shown in Fig.~\ref{Fig:gap-sol}. The upper part of the figure shows what happens if the solution obtained with DHT is used as an initial propagator in the solver that computes the convolutions using adaptive integration routines on a grid with 256 momentum values. We see that $\bar M^2_{\phi=0}(K)$ obtained in the first iteration deviates by $5\%-8\%$ from the solution obtained with DHT as a result of the fact that, as anticipated, the setting-sun calculated with DHT is not accurate. As the iteration progresses, $\bar M^2_{\phi=0}(K)$ departs even more from the used initial function and hence the converged solution is substantially different than the one obtained with DHT.

In the lower part of Fig.~\ref{Fig:gap-sol} we show $\bar M^2_{\phi=0}(K_{\rm min})$ as function of $|m_0^2|$ at four values of $g_0$. The difference between the curves obtained with the two ways of treating the convolution increases with the value of the coupling. This is due to the fact that the numerical error made in computing the convolution with DHT is magnified when the setting-sun is multiplied with a larger coupling. More importantly, the shape of the curves is compatible with the fact that the solution of \eqref{Eq:gap_eq_order_g2} is lost at some value of $m_0^2$ where $\bar M^2_{\phi=0}(K_{\rm min})$ is still finite. As a result the critical line cannot be determined.


\end{document}